\documentclass[12pt]{article}
\usepackage{amsmath}
\usepackage{amsfonts}
\usepackage{amssymb}
\usepackage{mathrsfs}
\usepackage{multicol}
\usepackage{color}
\usepackage{graphicx}
\usepackage{pstricks}
\usepackage{pst-node}
\usepackage{wrapfig}
\usepackage{enumerate}
\usepackage{caption}
\usepackage{cite}
\usepackage{subfigure}
\usepackage{amsfonts,amssymb}
\usepackage{amsthm}
\usepackage{amscd}
\definecolor{gray1}{gray}{0.1}
\definecolor{gray2}{gray}{0.2}
\definecolor{gray3}{gray}{0.3}
\definecolor{gray4}{gray}{0.4}
\definecolor{gray5}{gray}{0.5}
\definecolor{gray6}{gray}{0.6}
\definecolor{gray7}{gray}{0.7}
\definecolor{gray8}{gray}{0.8}
\definecolor{gray9}{gray}{0.9}

\newpsobject{showgrid}{psgrid}{subgriddiv=1,griddots=10,gridlabels=6pt}

\definecolor{dark-green}{rgb}{0,0.7,0}
\definecolor{dark-blue}{rgb}{0,0.2,0.5}
\definecolor{med-blue}{rgb}{0,0.7,1}
\definecolor{mblue}{rgb}{0,0.2,1}
\definecolor{cnc}{rgb}{0.8,0,0}
\definecolor{light-red}{rgb}{1,0.8,0.8}
\definecolor{dark-yelow}{rgb}{1,0.8,0}
\definecolor{light-blue}{rgb}{0.8,0.9,1}
\definecolor{verylight-blue}{rgb}{0.93,0.95,1}
\definecolor{light-yelow}{rgb}{1,0.9,0.8}
\definecolor{grey}{gray}{0.88}

\newpsobject{showgrid}{psgrid}{subgriddiv=1,griddots=10,gridlabels=6pt}

\newcommand{\ee}{\end{equation}}
\newcommand{\be}{\begin{equation}}

\begin{document}

\thispagestyle{empty}

\setlength{\abovecaptionskip}{10pt}

\begin{center}
{\Large\bfseries\sffamily{Scalarization 
of asymptotically Anti-de Sitter black holes
with applications to
holographic phase transitions}}
\end{center}
\vskip 3mm
\begin{center}
{\bfseries{\sffamily{Yves Brihaye$^{\rm 1}$, Betti Hartmann$^{\rm 2,3}$, Nath\'alia Pio Aprile$^{\rm 2}$, Jon Urrestilla$^{\rm 4}$ }}\\
\vskip 3mm{$^{\rm 1}$\normalsize Physique-Math\'ematique, Universit\'e de Mons-Hainaut, 7000 Mons, Belgium}\\
{$^{\rm 2}$\normalsize{Instituto de F\'isica de S\~ao Carlos (IFSC), Universidade de S\~ao Paulo (USP), CP 369, \\
13560-970 , S\~ao Carlos, SP, Brazil}}\\
{$^{\rm 2}$\normalsize{Department of Theoretical Physics, University of the Basque Country UPV/EHU, 48080 Bilbao, Spain}}}
\end{center}

\begin{abstract} 
We study the spontaneous scalarization of spherically symmetric, static and asymptotically Anti-de Sitter (aAdS) black holes in a scalar-tensor gravity model with non-mininal coupling of the form $\phi^2\left(\alpha{\cal R} + \gamma {\cal G}\right)$, where $\alpha$ and $\gamma$ are constants, while ${\cal R}$ and ${\cal G}$ are the Ricci scalar
and Gauss-Bonnet term, respectively.  Since these terms act as an effective ``mass'' for the scalar field,
non-trivial values of the scalar field in the black hole space-time are possible for  {\it a priori} vanishing scalar field mass. In particular, we demonstrate that the scalarization of an aAdS black hole requires the curvature invariant
$-\left(\alpha{\cal R} + \gamma {\cal G}\right)$ to drop below the Breitenlohner-Freedman bound close to the black hole horizon, while it asymptotes to a value well above the bound. 
The dimension of the dual operator on the AdS boundary depends
on the parameters $\alpha$ and $\gamma$ and we demonstrate that -- for fixed operator dimension -- the expectation value of this dual operator increases with decreasing temperature of the black hole, i.e. of the dual field theory.
When taking backreaction of the space-time into account, we find that the scalarization of the black hole
is the dual description of a phase transition in a strongly
coupled quantum system, i.e. corresponds to a holographic phase transition. A possible application are liquid-gas quantum phase transitions, e.g. in $^4$He. 
Finally, we demonstrate that extremal black holes with $AdS_2\times S^2$ near-horizon geometry {\it cannot support
regular scalar fields on the horizon} in the scalar-tensor model studied here. 

\end{abstract}

%%%%%%%%%%%%%%%%%%%%%%%%%%%%%%%%%%%%%%%%%%%%%%%%%%%%%%%%%%%%%%%%%%%%%%%%%%%%%%%
\section{Introduction}
With the advent of high precision, multi-messenger observations of black holes in a broad interval
of masses and sizes (see e.g. \cite{LIGO,EHT,CHANDRA}), it will soon be possible to test theoretical predictions of General Relativity
related to these objects. One of the most interesting questions is whether
classical black holes are, indeed, very simple, structureless objects that can be described by
a very small number of parameter -- the mass, charge and angular momentum -- and as such
fulfill the {\it No hair conjecture}. Theoretical black holes in asymptotically flat, 4-dimensional
Einstein-Maxwell theory have been proven to possess this feature \cite{nohairtheorems}, but
the existence of black holes with primary \cite{AGK,GKW} and secondary hair \cite{hairy_BHs}
in models with non-linear matter sources demonstrates clearly that it is far from obvious that
this conjecture holds true in other settings. 
In fact, while a number of theorems for scalar fields in black hole space-times exist \cite{no_scalar_hair}, black holes
can carry non-trivial scalar fields on their horizon when they are e.g. non-minimally coupled to the curvature of the space-time. This has been imvestigated extensively over the past years in the context of Horndeski scalar-tensor gravity models
 \cite{horndeski,Deffayet:2013lga,Charmousis:2014mia}. Indeed, in these models, 
 static, asymptotically flat black holes that carry scalar hair can be constructed \cite{Sotiriou:2014pfa, Babichev:2017ab}.  
 ``Spontaneous scalarization'' is a phenomenon that appears typically in models that contain non-minimal coupling terms of the form $f(\phi){\cal I}(g_{\mu\nu}, \Sigma)$, where $f(\phi)$ is a function of the scalar field, while ${\cal I}$ depends on the metric  $g_{\mu\nu}$ and/or other fields $\Sigma$. The scalar field then gets ``sourced''  by ${\cal I}$  and spontaneously scalarized black holes can be constructed for sufficiently large couplings. Recent examples include
the scalarization of static, uncharged, asymptotically flat black holes using ${\cal I}={\cal G}$, where ${\cal G}$ is the Gauss-Bonnet term and $f(\phi)=\phi^2$ \cite{Silva:2017uqg},
 different other forms of $f(\phi)$ with a single tem in $f(\phi)$ \cite{Doneva:2017bvd, Antoniou:2017acq, 
 Antoniou:2017hxj} or a combination of different powers of $\phi$ \cite{Minamitsuji:2018xde,Brihaye:2018grv}.
These studies have been extended to include charge \cite{Doneva:2018rou,Herdeiro:2018wub,Brihaye:2019kvj} as well
as a positive cosmological constant \cite{Bakopoulos:2018nui,Brihaye:2019gla,Bakopoulos:2019tvc}.

Modifications of General Relativity are usually motivated by assuming General Relativity to be only a 
classical, low energy limit of a (more general) Quantum Theory of gravity that should be applicable as well
at and close to the Planck scale. One of the best candidates for such a theory remains String Theory. One remarkable
prediction of String Theory is the so-called gauge/gravity duality, a conjecture that relates
gravity theories in $(d+1)$ space-time dimensions to gauge theories in $d$ dimensions \cite{ggdual}. The best tested and well studied
example is the Anti-de Sitter/Conformal Field Theory (AdS/CFT) correspondence \cite{adscft}, which
connects a gravity theory in $(d+1)$-dimensional AdS space-time to an SU(N) gauge theory on the $d$-dimensional
boundary of AdS. This duality is a weak-strong coupling duality such that ``weakly coupled'', classical gravity theories
in AdS can be used to describe strongly coupled quantum systems on the conformal boundary of AdS.
With a black hole present in the bulk of AdS, the quantum system can be studied at a given temperature.
Holographic phase transitions typically appear when lowering the black hole temperature and correspond 
to non-trivial matter fields forming on the black hole below a certain critical temperature \cite{Gubser:2008px}.
These ideas have mainly been applied to the description of high-temperature superconductivity in the framework
of holographic superconductors \cite{hhh,reviews}
as well as the description of the quark-gluon plasma (see e.g. \cite{Ammon:2015wua} and references therein).
The first studies have typically been conducted using scalar fields. In fact, as shown in \cite{BF}, a scalar field
becomes unstable in AdS if its mass $m$ drops below the so-called Breitenlohner-Freedman (BF) bound,
i.e. if $m^2 \leq m_{\rm BF}^2$. This bound is dimension-dependent and can be shown to lead
to formation of non-trivial scalar hair in a number of settings within asymptotically AdS (aAdS) space-time.

In this paper, we discuss a scalar-tensor gravity model in $(3+1)$-dimensional aAdS space-time. 
In this model, the scalar field is {\it a priori} massless, but couples non-trivially to the Ricci scalar
and Gauss-Bonnet term, respectively, of the space-time. As we will demonstrate in the following,
this non-trivial coupling corresponds to an ``effective mass'' for the scalar field and generates the formation of
non-trivial scalar hair on the black hole for sufficiently low temperature of the latter. When taking backreaction
of the space-time into account, this ``spontaneous scalarization'' can be interpreted as a holographic phase transition, i.e. a phase transition on a spatially
2-dimensional surface with order parameter given by a real-valued scalar field and appearing in a strongly coupled quantum system. 
Note that uncharged, static, spherically symmetric, aAdS black holes in a scalar-Gauss-Bonnet model
have been studied briefly in \cite{Bakopoulos:2018nui}, however, not with the emphasis on solutions
that possses a power law fall-off on the AdS boundary, a requirement that will be crucial for us in the following.

Our paper is organized as follows~: in Section 2, we discuss the model, while Section 3 deals with the probe limit, i.e.
the limit of vanishing backreaction of the space-time. In Section 4, the inclusion of backreaction is presented,
while Section 5 contains our conclusions.

%%%%%%%%%%
\section{The model}
%%%%%%%%%% 
In this paper, we study a scalar-tensor gravity model with the following action
\begin{equation}
\label{action}
S=\int {\rm d}^4x  \ \sqrt{-g} \left[\frac{{\cal R}}{2} - \Lambda +  \phi^2 \left(\alpha {\cal R} + \gamma {\cal G}\right)  -  \partial_{\mu} \phi \partial^{\mu} \phi  - \frac{1}{4}
 F_{\mu\nu } F^{\mu\nu}  \right]  \ ,
\end{equation}
where ${\cal R}$ is the Ricci scalar, ${\cal G}$ the Gauss-Bonnet term, $\Lambda < 0$ the cosmological constant, $F_{\mu\nu}=\partial_{\mu} A_{\nu} - \partial_{\nu}A_{\mu}$ the field stength tensor of a U(1) gauge field $A_{\mu}$ and $\phi$ a real-valued, massless scalar field that
is coupled to the Ricci scalar ${\cal R}$ as well as the Gauss-Bonnet term ${\cal G}$ given by
\begin{equation}
{\cal G} = (R^{\mu \nu \rho \sigma} R_{\mu \nu \rho \sigma} - 4 R^{\mu \nu} R_{\mu \nu} + R^2 ) 
\end{equation}
via the couplings $\alpha$ and $\gamma$, respectively. 
Variation of the action with respect to the metric, scalar field and U(1) gauge field leads to a set of coupled differential
equations that have to be solved numerically given appropriate boundary conditions.
Variation with respect to the scalar field, U(1) gauge field and metric, respectively, leads to the following set of coupled, non-linear differential equations~:
\begin{equation}
\label{eq:scalar}
\square\phi+\left(\alpha {\cal R} + \gamma {\cal G}\right) \phi =0   \ \ , \ \
\frac{1}{\sqrt{-g}}\partial_{\mu} \left(\sqrt{-g} F^{\mu\nu}\right)=0   \ \ , \ \
G_{\mu\nu} + \Lambda g_{\mu\nu} = T^{(\phi)}_{\mu\nu} + T^{({\rm EM})}_{\mu\nu} ,
\end{equation}
where
\begin{eqnarray}
T^{(\phi)}_{\mu\nu}&=&\partial_{\mu}\phi \partial_{\nu}\phi - \frac{1}{2} g_{\mu\nu} \partial_{\sigma}\phi \partial^{\sigma}\phi + 4\alpha \left[D_{\mu}(\phi \partial_{\nu} \phi) - g_{\mu\nu} D_{\sigma} (\phi \partial^{\sigma} \phi)\right] \nonumber \\
&-& 2\gamma \left(g_{\rho\mu} g_{\lambda\nu} + g_{\lambda\mu} g_{\rho\nu}\right)
\eta^{\kappa\lambda\beta\delta}\eta^{\rho\iota\sigma\tau} R_{\sigma\tau\beta\delta} D_{\iota} (\phi\partial_{\kappa}\phi)  \ 
\end{eqnarray}
and 
\begin{equation}
T^{({\rm EM})}_{\mu\nu}= F_{\mu\alpha}F_{\nu}^{\alpha} - 
\frac{1}{4} F_{\alpha\beta} F^{\alpha\beta}  \ .
\end{equation}
In the following, we will consider spherically symmetric, static black hole solutions. The Ansatz for the metric reads~:
\be
     ds^2 = - N \sigma^2 dt^2 + \frac{1}{N} dr^2 + r^2 \left(d\theta^2 + \sin^2\theta d\varphi^2\right)   
\ee
with $N\equiv N(r)$ and $\sigma\equiv \sigma(r)$ depending on the radial coordinate $r$ only. The explicit form
of the curvature invariants is then~:
\begin{equation}
\label{eq:ricci}
{\cal R}=\frac{1}{r^2\sigma}\left( - N'' r^2 \sigma - 3 N' \sigma' r^2 - 4 N'r \sigma
- 2 N \sigma'' r^2  - 4N\sigma' r 
- 2 N\sigma + 2\sigma\right)  \ ,
\end{equation}
\begin{equation}
\label{eq:GB}
{\cal G}=\frac{4}{r^2\sigma}\left(N'' N \sigma - N''\sigma + (N')^2\sigma + 
5N'\sigma' N - 3N'\sigma' + 2\sigma'' N^2 \
 - 2\sigma'' N\right)  \ .
\end{equation}
Assuming the symmetries
of the U(1) gauge field and the scalar field, respectively, to be equivalent to those of the space-time, we choose
$A_{\mu}dx^{\mu}\equiv V(r)dt$ and $\phi\equiv \phi(r)$. Inserting the Ans\"atze into the equations of motions leads
to a coupled set of non-linear ordinary differential equations of the form
\begin{eqnarray}
\label{eqs}
N' &=& {\cal F}_1(N,\sigma, V, V', \phi, \phi') \ \ , \ \
\sigma'= {\cal F}_2(N,\sigma, V, V', \phi, \phi') \ \ , \nonumber \\
V''&=& {\cal F}_3(N,\sigma, V, V', \phi, \phi') \ \ , \ \
\phi''= {\cal F}_4(N,\sigma, V, V', \phi, \phi') \ \ , \ \
\end{eqnarray}
where the prime now and in the following denotes the derivative with respect to $r$ and ${\cal F}_i$, $i=1,2,3,4$ are functions of the arguments.  The system (\ref{eqs}) has to be solved according to appropriate boundary conditions. At the regular horizon $r=r_h$ with $N(r=r_h)=0$ the condition for the scalar field reads~:
\be
\label{condition_horizon}
\left(\frac{\phi'}{\phi}\right)_{r=r_h}  = 
-\left(\frac{\alpha {\cal R} + \gamma {\cal G}}{N'}\right)_{r=r_h}  \ ,
\ee
while for the U(1) gauge field we have $V(r=r_h)=0$. As long as ${\cal R}$ and ${\cal G}$ are well behaved on the horizon, we would expect $\phi$ and $\phi'$ to be also well behaved there except when $N'$ becomes zero. This corresponds to the
extremal limit, where the Hawking temperature of the black hole given by
\begin{equation}
\label{temperature}
T_{\rm H} = \frac{1}{2\pi} \left(\sigma N'\right)\vert_{r=r_h} 
\end{equation}
tends to zero. For $r\rightarrow \infty$, we assume the space-time to be asymptotically AdS and the black hole to possess an electric charge $Q$, i.e. 
\begin{equation} 
\sigma(r\rightarrow \infty)\rightarrow 1 \ \ , \ \ 
N(r\rightarrow \infty)\rightarrow 1- \frac{\Lambda}{3} r^2  \  \ , \ \
V(r\rightarrow \infty)\rightarrow \frac{Q}{r_h} - \frac{Q}{r} \equiv \mu - \frac{Q}{r} \ ,
\end{equation}
where $\mu$ corresponds to the value of the U(1) gauge field on the conformal boundary $r\rightarrow \infty$
and can be interpreted as chemical potential in gauge/gravity applications. 
Finally, for $\Lambda < 0$, the scalar field has a power-law fall-off of the form~:
\begin{equation}
\label{eq:behaviour_boundary}
\phi(r\rightarrow\infty) \rightarrow \frac{\phi_{+}}{r^{\lambda_+}} + \frac{\phi_{-}}{r^{\lambda_-}}
\end{equation}
with
\begin{equation}
\label{eq:lambda_delta}
\lambda_{\pm} = \frac{3 \pm \sqrt{\Delta}}{2} \ \ , \ \ \Delta \equiv 9 + 48 \alpha +  32 \gamma \Lambda  \ ,
\end{equation}
i.e. the couplings $\alpha$ and $\gamma$ together with $\Lambda$ determine the dimension of the dual operator
on the conformal boundary in gauge/gravity  applications and $\phi_{\pm}$ can be interpreted as the expectation value
of this operator. Note that $\Lambda=0$, i.e. the asymptotically flat case,  is explicitly excluded here and is not a ``smooth limit'' of (\ref{eq:lambda_delta}). In fact, as shown in  \cite{Doneva:2017bvd, Silva:2017uqg,  Brihaye:2019kvj}
the scalar field always falls off like $\phi(r\rightarrow\infty)\sim \phi_0/r$ for $\Lambda=0$ (independent of $\alpha$ and $\gamma$), where $\phi_0$ is interpreted as the scalar charge of the solution. This interpretation is no longer possible in aAdS and we will hence in the following refer to $\phi_{\pm}$ as ``the expectation value of the dual operator on the conformal boundary'' -- using gauge/gravity terminology. Note that although our calculation is done
in Schwarzschild-like coordinates, the result for the power of
the scalar-field fall-off agrees with that obtained in the Fefferman-Graham construction, see Appendix B.

Let us mention as well that there is a scaling symmetry in the model, which reads (including all parameters and field)~:
\begin{eqnarray}
\label{scalingk1}
& & r\rightarrow \beta r \  , \  t\rightarrow t \  ,  \ M\rightarrow \beta M \ , \ Q\rightarrow \beta Q \  ,  \ 
\Lambda \rightarrow \frac{\Lambda}{\beta^2}  \ ,  \nonumber \\
& &    \ \alpha \rightarrow \alpha \  ,  \ \gamma\rightarrow \beta^2 \gamma \  ,  \  \phi_{\pm} \rightarrow  \beta^{\lambda_{\pm}} \phi_{\pm} \ \ ,  \ \ \mu\rightarrow \beta \mu \ ,
\end{eqnarray}
which scales the metric by $\beta^2$ and $A_{\mu}dx^{\mu}$ by $\beta$. 
This allows to set one parameter to a fixed value without loss of generality. In our numerical construction (see below),
we will often fix the horizon radius to $r_h=1$.

\section{The probe limit}

For vanishing scalar field $\phi(r)\equiv 0$, the model has explicit solutions~: the Schwarzschild-Anti-de Sitter (SAdS) solution for vanishing electric charge and the Reissner-Nordstr\"om-Anti-de Sitter (RNAdS) solution for non-vanishing electric charge, respectively. 
These solutions read~:
\begin{equation}
N(r)=1-\frac{2M}{r}+ \frac{Q^2}{2r^2} - \frac{\Lambda}{3} r^2 \ \ , \ \  \sigma\equiv 1 \ \ , \ \  V(r)= Q\left(\frac{1}{r_h} - \frac{1}{r}\right) \equiv \mu - \frac{Q}{r}  \ .
\end{equation}
$M$ is the mass and $Q$ the electric charge of the solution. The event horizon 
$r_h$ is the largest root of the equation $N(r_h)=0$ and leads to the following relations
\begin{equation}
M=\frac{1}{2}r_h -  \frac{\Lambda}{6} r_h^3 + \frac{Q^2}{4 r_h} \ \ , \ \ 
\Lambda=3\left(\frac{1}{r_h^2} - \frac{2M}{r_h^3} + \frac{Q^2}{2 r_h^4}\right) \ \ , \ \ 
Q=\pm \sqrt{ - 2 r_h^2 + \frac{2}{3}\Lambda r_h^4 + 4 Mr_h} \ .
\end{equation}
The Hawking temperature  (\ref{temperature})  of the RNAdS black hole reads
\begin{equation}
\label{eq:temperature_RNAdS}
2\pi T_{\rm H}= - \Lambda r_h - \frac{Q^2}{2 r_h^3}+ \frac{1}{r_h}
\end{equation}
and becomes zero for the extremal solution, which fulfills $N(r_{h,ex})=N'(r_{h,ex})=0$. This latter condition gives~:
\begin{equation}
r_{h,ex}=\sqrt{\frac{1-\sqrt{1-2\Lambda Q_{\rm ex}^2}}{2\Lambda}} \  \ , \ \  Q_{\rm ex}=\pm r_{h,ex} \sqrt{2\left(1-\Lambda r_{h,ex}^2\right)} \ .
\end{equation}
Hence, the extremal possible charge $Q_{\rm ex}$ increases (decreases) from $+\sqrt{2} r_{h,ex}$  (from $-\sqrt{2} r_{h,ex}$) at $\Lambda=0$ when decreasing $\Lambda$ , which is related to the additional attractive nature of the negative cosmological constant. 

In the following, we will assume this space-time background to be fixed and will consider a scalar field that does not 
backreact onto the space-time. The linear scalar field equation then reads~:
\be
\label{background_equation}
     \frac{1}{r^2}\left(r^2 N \phi'\right)' +\left(\alpha {\cal R} + \gamma {\cal G}\right)\phi = 0 \ ,
\ee
where the explicit form of the  Gauss-Bonnet term and the Ricci scalar, respectively, for the RNAdS solution are~:
\begin{eqnarray}
    {\cal G} = \frac{8}{3} \Lambda^2 + \frac{2}{r^8} \left( 24 r^2 M^2 - 24 r M Q^2 + 5 Q^4\right) \ \ , \ \
    {\cal R}= 4\Lambda \ .
\end{eqnarray}
Moreover, as is obvious
from these expressions, both ${\cal R}$ and ${\cal G}$ are constant on the AdS boundary at $r\rightarrow \infty$.
Hence, all arguments related to the holographic interpretation, regularization and renormalization of a scalar field
model with massive scalar field in asymptotically AdS apply also here (see e.g. \cite{Skenderis:2002wp} for a discussion on these issues).

\subsection{The Breitenlohner-Freedman bound and parameter restrictions}
\label{subsec:BF}

The $(d+1)$-dimensional asymptotically AdS (aAdS) space-time with AdS radius $\ell=\sqrt{-d(d-1)/(2\Lambda)}$  possesses a classical instability for 
a massive, real scalar field with equation $(\square - m^2)\phi=0$ if the mass $m$ is below the so-called  Breitenlohner-Freedman (BF) bound \cite{BF}, i.e. for $m^2 < m^2_{\rm BF}=-d^2/(4\ell^2)$, a non-trivial scalar field will form.
In our case, the ``mass'' is given by the term $m^2\equiv m^2_{\rm eff}=-(\alpha{\cal R}+\gamma{\cal G})$, i.e.
we would expect $(d+1)$-dimensional aAdS to form a non-trivial scalar for
\begin{equation}
\label{eq:BF}
\alpha{\cal R}+\gamma{\cal G} = 4\alpha \Lambda + \gamma{\cal G} \geq \frac{d^2}{4\ell^2} = -\frac{d \Lambda}{2(d-1)}  \ .
\end{equation}

In the following, we will require that $(3+1)$-dimensional aAdS  is stable with respect to the formation of a scalar field, i.e. that asymptotically, the space-time is ``pure'' AdS. We hence obtain the following
restriction on the parameters when using the asymptotic forms of the curvature tensors
\begin{equation}
\label{eq:re1}
4\alpha \Lambda + \frac{8}{3} \gamma \Lambda^2 \ \leq \ \frac{9}{4\ell^2}   \ . 
\end{equation}
A quick inspection of (\ref{eq:lambda_delta}) demonstrates that the requirement $\Delta \geq 0$ is exactly
(\ref{eq:re1}).

On the other hand, we would like a non-trivial scalar field to form close to the black hole horizon. In the following,
we will demonstrate that (\ref{eq:BF}) is fulfilled close to the event horizon of the scalarized black holes that we present in this paper. Note that the scalarization of black holes for $\alpha > 0$, $\gamma=0$ is impossible
due to the above argument. 

\subsection{Numerical results}
Equation (\ref{background_equation}) has to be solved numerically since - to our knowledge - no analytical solutions to this equation exist for generic values of $\alpha$ and $\gamma$. We have solved the equations using an adaptive grid collocation solver \cite{COLSYS}. We have also chosen $\phi_-=0$ for all our calculations. Moreover, we can choose $\phi(r_h)=1$ as well as $r_h=1$ without loosing generality. The latter condition fixes the mass $M$ in terms of $Q$ and $\Lambda$.
In the following we will only discuss cases with $\Lambda > -3$ in order to ensure that the AdS radius $\ell=\sqrt{-\Lambda/3}$ is larger than
the horizon radius $r_h$.

\subsubsection{$\alpha=0$}
\begin{figure}[ht!]
\begin{center}
\input{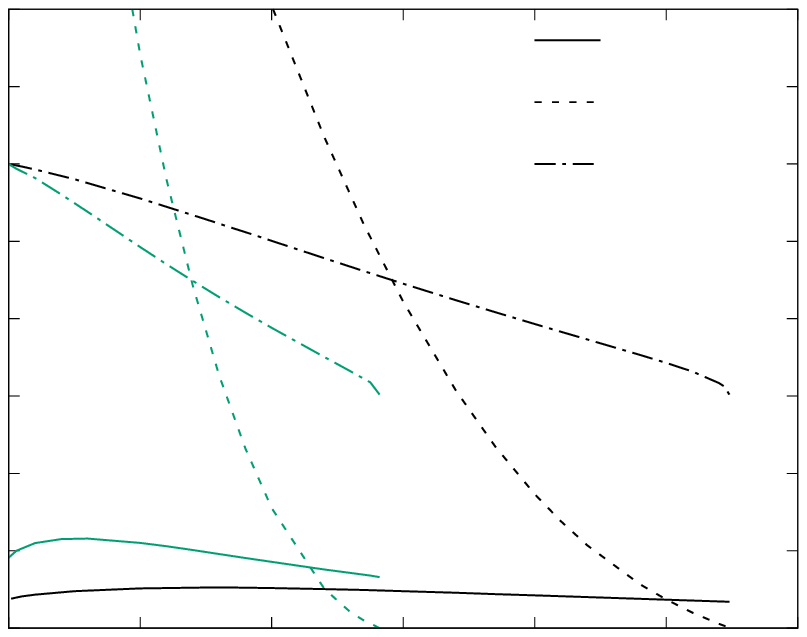}
\caption{We give the value of $\gamma$ (solid) for which non-trivial solutions of the scalar field equation 
(\ref{background_equation}) exist in dependence on $-\Lambda/3\equiv 1/\ell^2$ for a SAdS background ($Q=0$, black)
and a RNAdS background with $Q=1$ (green). We also give $\Delta$ (dashed) as well as the value of $\lambda_+$ (dotted-dashed). The latter corresponds to the dimension of the operator on the conformal boundary in the gauge/gravity duality interpretation. Note that all curves stop at a given value of $\Lambda$ because $\Delta\rightarrow 0$ there.} 
\label{fig1}
\end{center}
\end{figure}

In the limit $Q=0$, the black hole is given by the Schwarzschild-AdS (SAdS) solution which has 
vanishing chemical potential $\mu=0$.  As is well known from the $\Lambda=0$ case, spontaneous scalarization
of the Schwarzschild black hole appears for a very specific value of $\gamma=\gamma_{\rm cr}$, in our choice of couplings and prefactors
for $\gamma_{\rm cr}(\Lambda=0)\approx 0.18$ \cite{Silva:2017uqg}.
We would hence expect something similar to appear in an asymptotically AdS space-time. 
Indeed, we find that spontaneous scalarization appears for $\gamma = \gamma_{\rm cr}(\Lambda)$. We show the
dependence of $\gamma_{\rm cr}$ on $\Lambda$ in Fig. \ref{fig1} (black solid curve). Decreasing $\Lambda$, $\gamma_{\rm cr}$ first increases slightly and then decreases again, but shows little dependence on the value of $\Lambda$.
This means that the value of the curvature radius of the space-time, i.e. the AdS radius, has little influence on the
scalarization process of the black hole as long as this radius
is (much) larger than the horizon radius of the black hole. This seems reasonable since the scalarization happens on and close to the horizon.
$\Lambda$ can only be decreased to a minimal value of $\Lambda_{\rm *} \approx -1.65$ corresponding to an AdS radius
of $\ell\approx 1.35$ which is comparable in size to the horizon radius $r_h=1.0$. At this value of $\Lambda$, the power-law fall off of the scalar field
function is no longer possible since $\Delta$ (black dashed) tends to zero. Inserting $\Lambda_{\rm *}$ into the equation $\Delta=0$ gives $\gamma_{\rm *}\approx  0.17$, which is in perfect agreement with our numerical results.   
Accordingly, the power of the scalar
field fall-off $\lambda_+$ (equivalent to the dimension of the dual operator on the conformal boundary, black dotted-dashed) ranges from three at $\Lambda=0$ to $3/2$ at $\Lambda_{\rm *}$. \\

The case for $\Lambda=0$ and $Q\neq 0$  has been studied in \cite{Brihaye:2019kvj}.  In particular, it was found that two independent and disjoined branches
of charged, spontaneously scalarized black holes exist: one is the solution that tends to the uncharged solution
of \cite{Doneva:2017bvd, Silva:2017uqg, Antoniou:2017acq} and exists for $\gamma_{\rm cr} > 0$, while the second 
branch appears close to extremality of the RN solution and requires $\gamma_{\rm cr} < 0$. 
In order to understand the pattern, we have first fixed $Q=1$ and compared the results with those for $Q=0$, see Fig. \ref{fig1}.  Qualitatively, the results are very similar to those of the uncharged case when considering only $\gamma_{\rm cr} > 0$. Quantitatively, we observe that $\gamma_{\rm cr}$ (solid green) is always larger than in the uncharged case and that the
value of $\Lambda_{\rm *}$ at which $\Delta \rightarrow 0$ increases with charge. For $Q=1$ we find 
$\Lambda_{\rm *}\approx -0.84$, which -- using $\Delta=0$ -- gives the value $\gamma_{\rm *}\approx 0.33$, again in excellent agreement with our numerics (see green dashed curve). We observe that again $\lambda_+\in [\frac{3}{2}:3]$ (green dotted-dashed curve), but that this variation is related to a smaller variation of $\Lambda$
as in the uncharged case. 

\begin{figure}[ht!]
\begin{center}
\input{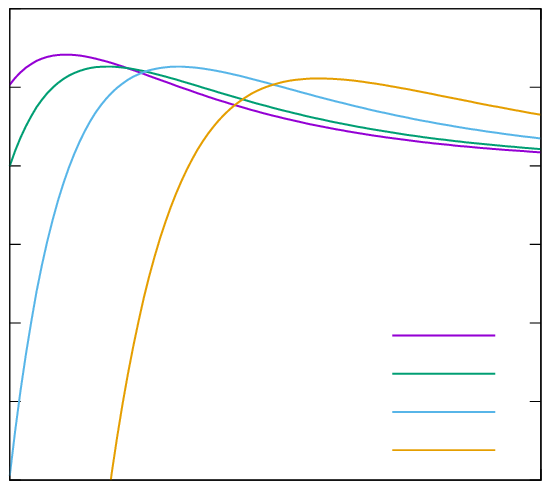}
\input{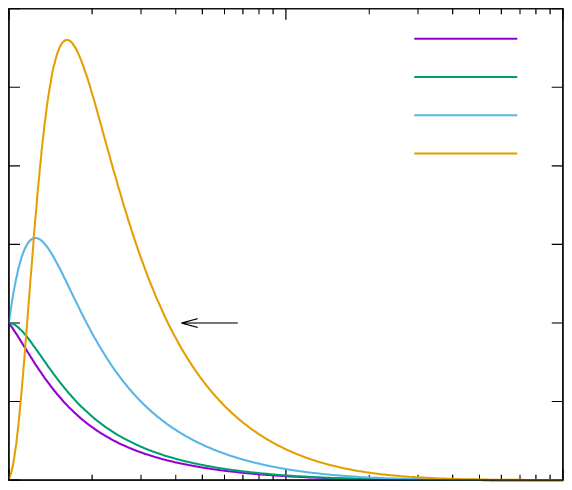}
\caption{We show the Gauss-Bonnet term $\gamma_{\rm cr}{\cal G}$ for the RNAdS solutions with $r_h=1$, $\Lambda=-0.006$ and different values of $Q^2$ (left) as well as the scalar field solution $\phi(r)$ in the corresponding background RNAdS space-time (right). }
\label{fig:GB}
\end{center}
\end{figure} 

In Fig. \ref{fig:GB} we show some typical solutions for $\Lambda=-0.006$ and different charges $Q$.  
Following the discussion in \ref{subsec:BF}, we  would expect $-\gamma_{\rm cr}{\cal G}$ to drop below
the BF bound close to the horizon of the black hole. For our choice of $\Lambda$, the value of BF bound is
 $m^2_{\rm BF}=-0.0045$. Inspection of Fig. \ref{fig:GB} (left) demonstrates that
 close to the horizon $r_h=1$, we find $-\gamma_{\rm cr} {\cal G} < -0.0045$, while
 asymptotically the requirement $\Delta \geq 0$ ensures stability of AdS. Correspondingly, non-trivial scalar
 fields appear close to the horizon, see Fig. \ref{fig:GB} (right). We also observe that the larger the charge $Q$, the 
 larger we have to choose $\gamma_{\rm cr}$ to find scalarized black holes. E.g. for
 the charges given in Fig. \ref{fig:GB} we find $\gamma_{\rm cr}=0.48$, $0.60$, $1.00$ and $2.00$ for
 $Q^2=0.9946$, $1.1032$, $1.3083$, $1.8106$, respectively.
To state it differently: charged black holes
 require a stronger scalar-tensor coupling in order to be scalarized as compared to their uncharged counterparts.
 
 When increasing $Q$, we observe a phenomenon that exists also for $\Lambda=0$ and was first discussed in
 \cite{Brihaye:2019kvj}~: the GB term becomes negative close to the black hole horizon due to the approach
 of extremality, i.e. the approach to a solution with near-horizon geometry $AdS_2\times S^2$. 
  For $\Lambda=-0.006$ and $r_h=1$, the value of the extremal charge is $Q^2_{\rm ex}=1.988$, but we
  see the appearance of a negative valued GB term already at $Q^2=1.3083$ and $Q^2=1.8106$.

\begin{figure}[ht!]
\begin{center}
\input{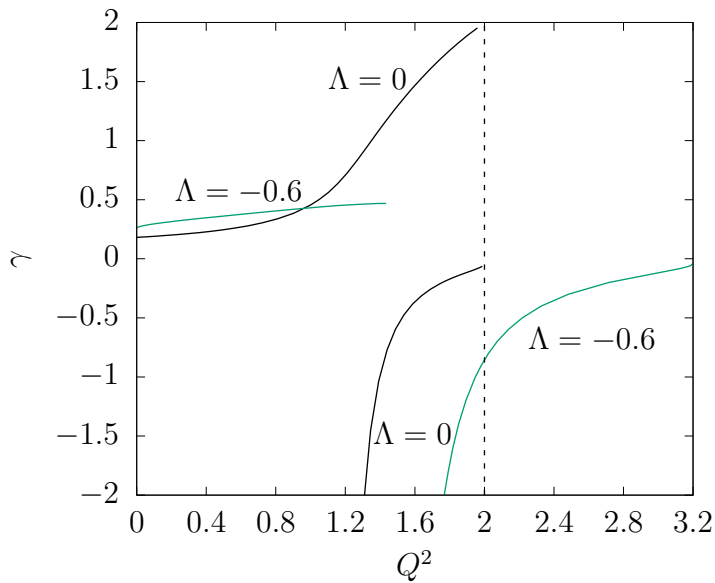}
\caption{We give the value of $\gamma_{\rm cr}$ for which non-trivial solutions of the scalar field equation 
(\ref{background_equation}) exist in dependence on $Q^2$ and for $\Lambda=0$ (black) and $\Lambda=-0.6$ (green), respectively. The dashed vertical line indicates the extremal limit $Q^2_{\rm ex}=2.0$ of the charged background solution for $\Lambda=0$, i.e.
the RN solution, while $Q^2_{\rm ex}=3.2$ for $\Lambda=-0.6$. }
\label{fig:positive_negative}
\end{center}
\end{figure}

To evaluate the influence of the cosmological constant on this phenomenon in more detail,
we have chosen $\Lambda=-0.6$ (with $Q^2_{\rm ex}=3.2$)  and compared the two $\gamma_{\rm cr}$ branches with those
present for $\Lambda=0$ (with $Q^2_{\rm ex}=2.0$). The results are shown in Fig. \ref{fig:positive_negative}.
We observe that while the positive $\gamma_{\rm cr}$ branches exists for $Q^2\in [0:Q^2_{\rm ex}]$ 
for $\Lambda=0$, this is no longer true for $\Lambda=-0.6$. The positive $\gamma_{\rm cr}$ branch stops
when $\Delta=0$, i.e. for $\gamma_{\rm cr}=0.46875$. We find the corresponding value of the charge to be $Q^2\approx 1.44$, well below the extremal charge $Q^2_{\rm ex}=3.2$. To state it differently, the requirement of asymptotically AdS space-time being stable, i.e. the BF bound, imposes a stronger restriction on the solutions -- at least for sufficiently small $\Lambda$ -- than the requirement for the existence of a black hole horizon.
This also leads to the observation that close to the extremal limit, aAdS black holes can only be scalarized for $\gamma_{\rm cr} < 0$. While for $\Lambda=0$ an interval of $Q^2$ exists, for which
black holes can be scalarized for positive and negative $\gamma_{\rm cr}$, this is no longer the case
for $\Lambda=-0.6$. In fact, we find that $Q^2\rightarrow 1.6$ for $\gamma\rightarrow -\infty$.
Hence, non-trivial scalar field solutions to (\ref{background_equation}) exist either for $\gamma$ positive or
$\gamma$ negative if $-\Lambda$ is sufficiently large.

\subsubsection{$\alpha \neq 0$}
\begin{figure}[ht!]
\begin{center}
\input{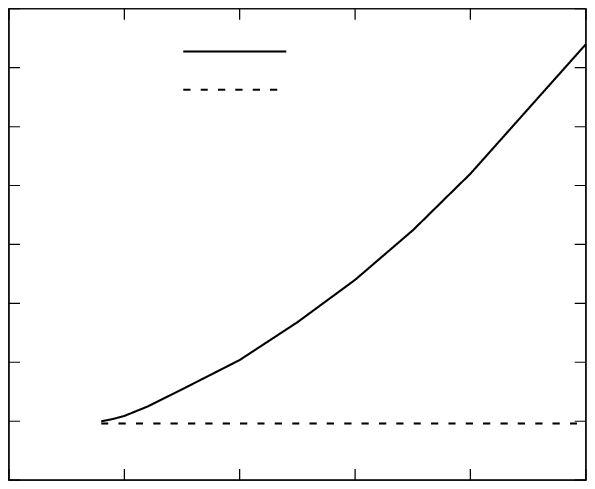}
\input{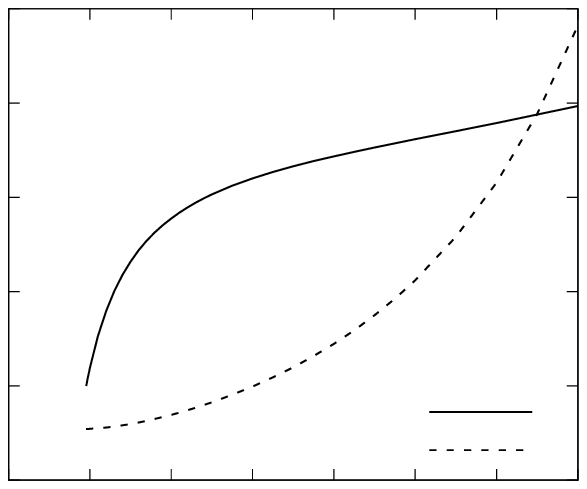}
\caption{We show the domain of existence of non-trivial solutions to (\ref{background_equation}) in the $\gamma$-$\alpha$-plane for $\Lambda=-0.006$ (left). We demonstrate the approach to $Q=0$ for $\alpha=0.0$ (right). Clearly, the derivative $\partial_r\phi(r_h)$ changes sign close to the approach, but stays finite.}
\label{fig:domain_ads}
\end{center}
\end{figure}

We have studied the case $\Lambda=-0.006$ and our results are shown in Fig. \ref{fig:domain_ads}, where
we present the domain of existence of scalarized RNAdS black holes in the $\gamma$-$\alpha$-plane.
This clearly shows that
solutions for $\gamma=0$ are not possible and that the $\alpha{\cal R}$ term always requires the presence of
the $\gamma{\cal G}$ term as well in order to achieve scalarization.
The domain of existence is limited by two phenomena~: $\Delta$ tending to zero, which gives a lower bound on $\alpha$
and $Q=0$, which gives an upper bound on $\alpha$, respectively.
From $\Delta=0$ using $\Lambda=-0.006$, we find $\alpha^{({\rm min})}_{\rm cr}=0.004\gamma - 0.1875$, hence
the $\Delta=0$ curve in Fig. \ref{fig:domain_ads} shows little dependence on $\gamma$. Since the presence
of the charge $Q$ leads to the presence of negative valued terms in ${\cal G}$, we would expect that
decreasing $Q$ allows to increase $\alpha$. This can be done until $Q=0$. In this limit, the requirement for
scalarized black holes to exist is (see (\ref{eq:re1})):
\begin{equation}
4\alpha \Lambda +\gamma \left(\frac{8}{3}\Lambda^2 + \frac{48 M^2}{r^6}\right) \geq -\frac{3\Lambda}{4}
\end{equation}
close to the black hole horizon. Using $r_h=1.0$, $\Lambda=-0.006$ this becomes
\begin{equation}
\label{eq:alpha}
\alpha \leq -0.01875 + \gamma\left(0.0004 + \frac{50.2002}{r^6}\right)  \ .
\end{equation}
Now, we would need this bound to be fulfilled somewhere outside and close to the horizon in order to observe
scalarization, i.e. the bound depends on the actual value of $r$ (or better: range of $r$), which can only be found numerically. But (\ref{eq:alpha}) clearly demonstrates that when increasing $\gamma$, we can increase $\alpha$, in agreement with our numerical results.

In summary, we can scalarize the SAdS solution and then increase the charge $Q$ up to the point where $\Delta=0$. 
In Fig. \ref{fig:domain_ads} (right), we show the approach to $Q=0$ for $\alpha=0.0$. We observe that close to $Q=0$, the value of the derivative of the scalar function on the horizon changes sign and becomes negative.

\subsubsection{Fixing the operator dimension}
In gauge/gravity duality applications it is often useful to fix the dimension of the dual operator. This corresponds to
fixing $\lambda_+$, i.e. fixing $\Delta$.  We can then express the Hawking temperature (\ref{eq:temperature_RNAdS}) in terms of $\lambda_+$ and $Q^2$ as follows
%%% lam=( - 12*alp + lamp**2 - 3*lamp)/(8*gam)
\begin{equation}
\label{eq:temperature_power_fixed}
2\pi T_{\rm H}=\frac{ \left[\left(3-\lambda_+\right)\lambda_+ + 12\alpha\right] r_h^4 + 4(2r_h^2 - Q^2)\gamma}{8\gamma r_h^3}
\end{equation}
where
\begin{equation}
\label{eq:rel_gamma_lam}
\gamma=-\frac{12\alpha + \lambda_+ (3-\lambda_+)}{8\Lambda} \ .
\end{equation}
We have investigated this case for $\alpha=0$, $\gamma > 0$ for which $\lambda_+$ can be chosen
to lie within the interval $\lambda_+\in [\frac{3}{2}:3[$. Note that we cannot reach $\lambda_+=3$, because
this would require either $\gamma=0$, in which the scalar field would always be trivial due to existing
no-hair theorems, or $\Lambda=0$, in which case the space-time is asymptotically flat and the scalar field
falls off like $\sim 1/r$.  

\begin{figure}[ht!]
\begin{center}
\input{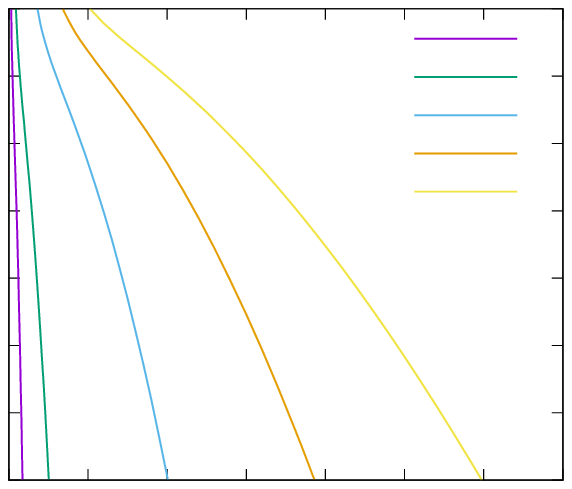}
\input{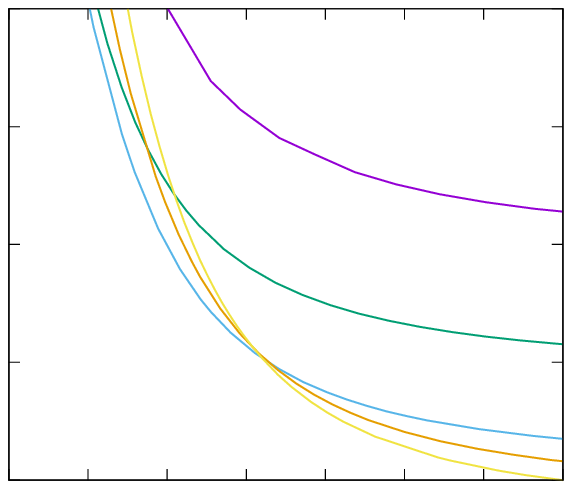}
\caption{We show the value of $Q^2$ in dependence on $-\Lambda/3=1/\ell^2$ for which non-trivial solutions to (\ref{background_equation}) exist for various values of fixed $\lambda_+$, i.e. dimension of the dual
operator on the conformal boundary, and $\alpha=0$ (left).  We also give the value of the expectation value of the dual operator
with dimension $\lambda_+$, $(\phi_+)^{1/\lambda_+}$, on the boundary in function of the Hawking temperature $T_{\rm H}$ (right, same colour coding as left).  }
\label{fig:power_fixed}
\end{center}
\end{figure}

Our results for various
fixed values within the given interval for $\lambda_+$ are shown in Fig. \ref{fig:power_fixed}. We find that $Q^2$ is a decreasing function of $-\Lambda$ (or equivalently decreasing AdS radius $\ell$) -- independent of the value of $\lambda_+$, see Fig.\ref{fig:power_fixed} (left). The largest possible $-\Lambda$ on each individual branch
is reached at the $Q=0$ solution, i.e. when the background is given by the Schwarzschild-Anti-de Sitter (SAdS) solution. To state it differently~: in order for charged black holes in aAdS to scalarize, we have to choose
the AdS radius larger (as compared to the horizon radius) than for an uncharged aAdS black hole. The larger
the charge $Q$, the larger we have to choose $\ell$ in order to achieve scalarization.

When decreasing $\lambda_+$, we find that for a fixed value of the charge $Q$, we have to increase $-\Lambda$ (decrease $\ell$) in order for the black hole to scalarize, i.e. the smaller the dimension of the dual
operator on the conformal boundary, the smaller we have to choose the AdS radius $\ell$ (in comparsion
to the horizon radius) in order to find non-trivial scalar fields. In Table \ref{table1}, we give the
values of $-\Lambda/3$ (or equivalently $\ell$) as well as the corresponding value of $\gamma$ (see (\ref{eq:rel_gamma_lam})) for which non-trivial solutions to (\ref{background_equation}) exist in the
$Q=0$ background in dependence on $\lambda_+$. For these values, the scalar field behaves 
according to (\ref{eq:behaviour_boundary}) for $\phi_-\equiv 0$. Using the terminology of the gauge/gravity duality, we can then interpret $\phi_+$ as the expectation value of the corresponding dual operator
with dimension $\lambda_+$.  In Fig.\ref{fig:power_fixed} (right) we show the value of $(\phi_+)^{1/\lambda_+}$ in dependence on
the Hawking temperature $T_{\rm H}$ of the black hole, which is equal to the temperature of the dual field theory on the AdS boundary. We find that for all fixed $\lambda_+$ that we have studied, $(\phi_+)^{1/\lambda_+}$ increases with decreasing temperature, which is equivalent to increasing charge $Q$. In fact, the branches shown here start at $Q=0$ (large $T_{\rm H}$) and end at the extremal solution with $T_{\rm H}=0$.  We find that -- except for $\lambda_+=2.97$, i.e.
a value of $\lambda_+$ close to the limiting value $\lambda_+=3$ -- the strongest increase of $(\phi_+)^{1/\lambda_+}$
happens approximately at the same $T_{\rm H}$. 

\begin{table}
\begin{center}
\begin{tabular}{|c||c|c|c|c|}
\hline
$\lambda_+$ & $-\Lambda/3$ & $\ell$ & $\gamma$ & $\phi_+$\\
\hline
$2.97$ & $0.0172 $ & $7.625$ & $0.216 $ & $9.380$\\
$2.90$ & $0.0504 $ & $4.454$ & $0.240$   & $3.535$\\
$2.50$ & $0.2000  $ & $2.236$ & $0.260 $ & $1.249$ \\
$2.00$ & $0.3860 $ & $1.610$ & $0.216 $ & $0.917$\\
$1.50$ & $0.5900 $ & $ 1.302 $ & $0.159 $ & $0.816$\\
\hline
\end{tabular}
\caption{Values of  $-\Lambda$, $\ell$ and $\gamma$ in dependence on $\lambda_+$ for
which non-trivial solutions to (\ref{background_equation}) exist in the background of a Schwarzschild-Anti-de Sitter (SAdS) solution, i.e. for $Q=0$, and for $\alpha=0$. We also give the corresponding value of the expectation value of
the dual operator on the boundary, $\phi_+$.}
\label{table1}
\end{center}
\end{table}

\section{Including backreaction}
\begin{figure}[ht!]
\begin{center}
\input{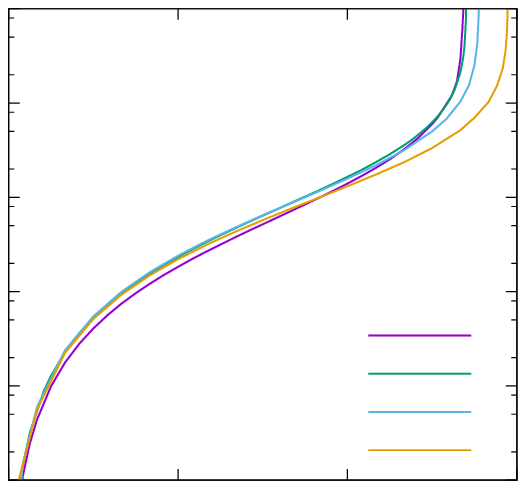}
\input{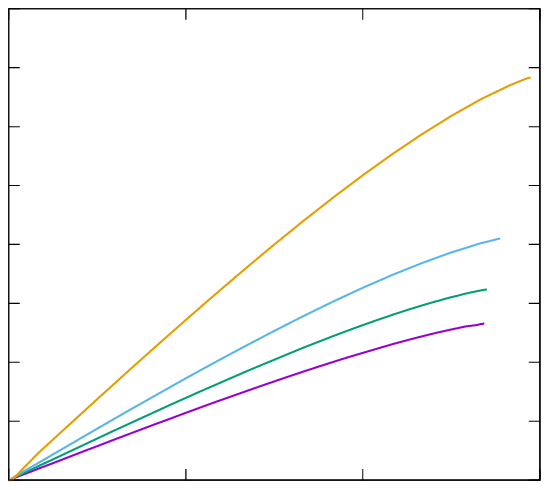}
\caption{We show the derivative of the metric function $\sigma(r)$ at the horizon, $\partial_r\sigma(r_h)$, in dependence on the value of the scalar field on the horizon, $\phi_h$, for spontaneously scalarized black holes including backreaction, $\gamma=0.3$ and different values of $\lambda_+$ (left).  We also give the value of the condensate $\phi_+$ in dependence on $\phi_h$ for these solutions (right, same colour coding as left).  }
\label{fig:backreaction1}
\end{center}
\end{figure} 

With the knowledge of the parameters for which non-trivial solutions to (\ref{background_equation}) exist, we can treat the full back-reacted problem and construct spontaneously scalarized black holes as solutions to the set of coupled, non-linear differential equations (\ref{eqs}). In fact, while $\phi\vert_{r=r_h}\equiv \phi_h$ is fixed in the linearized problem, it can now be varied continuously and is a free parameter. We have hence chosen the set of parameters $(\Lambda, \alpha, \gamma)$ for which a solution to (\ref{background_equation}) exists (again choosing $r_h=1$) and have numerically constructed a branch of solutions
characterized by $\phi_h$.  Our numerical results indicate that $\phi_h \in [0, \tilde{\phi}_h]$. While the Hawking temperature 
$T_{\rm H}$, the mass $M$ and the electric charge $Q$ depend very weakly on $\phi_h$, 
we observe that the derivative of the metric function $\sigma$  -- which itself is no longer trivially equal to unity -- at the horizon, $\partial_r\sigma\vert_{r=r_h}\equiv \partial_r\sigma(r_h)$, diverges when $\phi_h\rightarrow \tilde{\phi}_h$ indicating that the branch of spontaneously scalarized black holes tends to a singular solution. Equally, the Ricci scalar ${\cal R}$ (see (\ref{eq:ricci}))  and the Gauss-Bonnet term ${\cal G}$ (see \ref{eq:GB})) diverge on the horizon. 
Our numerical results indicate that we reach this singular solution before reaching the extremal limit
with $T_H=0$. The fact that black holes with strong curvature cannot be scalarized to give 
regular black holes with hair has recently been discussed in \cite{Bakopoulos:2019tvc} and our results
indicate that this is also true in aAdS.  To state it differently~: the non-minimal coupling between curvature and scalar field
does not allow to find an extremal, spontaneously scalarized black hole -- a statement that we prove explicitly in the Appendix. In fact, this is shown in Fig. \ref{fig:backreaction1} (left)
for $\alpha=0$, $\gamma=0.3$ and different fixed dimensions of the dual operator $\lambda_+$ (or equivalently
different values of $\Lambda$). In the limit $\phi_h\rightarrow 0$, the solution tends to the RNAdS solution with
$\phi(r)\equiv 0$ and $\sigma(r)\equiv 1$. Increasing $\phi_h$ from zero, the solution becomes a (only numerically known) charged black hole with scalar hair for which $\partial_r\sigma(r_h)$ increases. The value of $\phi_+$ also increases
with increasing $\phi_h$, see Fig. \ref{fig:backreaction1} (right). We find that the larger the $\lambda_+$, the larger $\phi_+$ for a given value of $\phi_h$. 

\subsection{Holographic phase transitions}
In the case without backreaction, the scalar field equation (\ref{background_equation}) is linear and hence
the value of the scalar field $\phi(r)$ has no physical meaning. In our model it is thence possible to describe holographic phase transitions only when including the backreaction of the metric. Note that
this is in contrast with the study of holographic superconductors \cite{hhh,reviews}, where
the scalar field equation is non-linear already for a fixed background metric due to a minimal coupling  to a U(1) gauge field
\cite{Gubser:2008px}.

\begin{figure}[ht!]
\begin{center}
\input{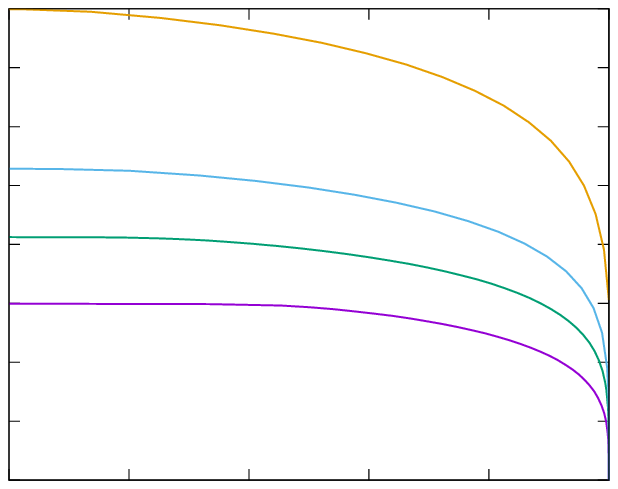}
\input{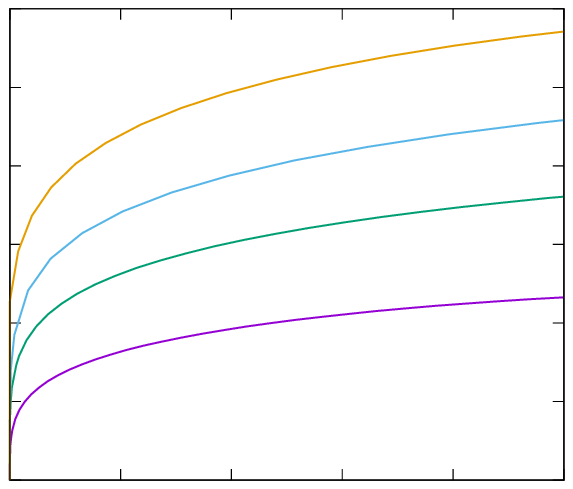}
\caption{We give the dimensionless condensate $\phi_+^{1/\lambda_+}/T_c$ ($\phi_+^{1/\lambda_+}/\mu_c$) 
in dependence on the temperature $T$ divided by the critical temperature $T_c$ (in dependence on the chemical potential $\mu$ divided by the critical chemical potential $\mu_c$) for scalarized black holes including backreaction with $\gamma=0.3$ and several values of $\lambda_+$ (left (right)).  }
\label{fig:backreaction2}
\end{center}
\end{figure}

When solving the full set of non-linear equations, we find that the model in the bulk can be interpreted
as the dual description of a phase transition in a strongly coupled quantum system that ``lives'' on the boundary of
AdS. This is shown in Fig. \ref{fig:backreaction2} (left), where we give the dimensionless
value of the condensate $\phi_+^{1/\lambda_+}/T_c$
in dependence of the temperature $T\equiv T_{\rm H}$, which we now interpret as the temperature of the dual
field theory and hence omit the index ``H''. Close to the critical temperature $T=T_c$, the condensate
curve shows the typical Ginzburg-Landau type behaviour $(\phi_+)^{1/\lambda_+}\sim \sqrt{1-T/T_c}$, which signals
a second order phase transition at $T=T_c$. Increasing the operator dimension $\lambda_+$, the value of the condensate
increases at a given temperature $T$. We have also studied the behaviour of the condensate 
$\phi_+^{1/\lambda_+}/\mu_c$ in dependence
of the value of the U(1) gauge field on the AdS boundary, i.e. the chemical potential $\mu$. Our results
are shown in Fig. \ref{fig:backreaction2} (right). For $\mu\geq \mu_c$, we find that non-trivial condensates exist
and, again, we find that increasing the operator dimension increases the value of the condensate for a fixed value of $\mu$.
The described phase transition, however, cannot be extended all the way down to $T=0$ because of the
reasons mentioned above. Rather, we find that below a given temperature (above a given chemical potential)
the value of the condensate becomes constant, i.e. is practically independent of the temperature (chemical potential), when
decreasing (increasing) it further.

Let us mention that in our numerical study here, we have fixed $r_h=1$, i.e. the entropy of the black hole ${\cal S}=\pi r_h^2$ to be constant.
In gauge/gravity applications it is often more interesting to study the case of fixed charge $Q$. Our data can easily be
transformed to this latter case by using the scaling symmetry (\ref{scalingk1}) that leaves $\lambda_+$ unchanged.  
The relevant quantities (denoted by a hat, $\hat{Q}=1$) then are~:
\begin{equation}
\hat{T}_{\rm H}= Q T_{\rm H} \ \ , \ \   \hat{\phi}_{\pm} = \frac{\phi_{\pm}}{Q^{\lambda_{\pm}}} \ \ , \ \ 
\hat{\gamma}=\frac{\gamma}{Q^2} \ \ , \ \   \hat{\mu}=\frac{\mu}{Q}
\end{equation}
where we have to exclude the $Q=0$ case. Since the uncharged case is of importance in our study
and scalarization happens due to the non-minimal coupling (in contrast to e.g. studies of
holographic superconductors, where the presence of the U(1) gauge field is essential for the minimally coupled
scalar field to form, see e.g. \cite{Gubser:2008px, hhh}), we have presented our results for fixed
entropy ${\cal S}$ of the black hole. 
\\

Due to the fact that the scalar field 
in our model is real, applications of our results are phase transitions
that involve real-valued order parameters rather than complex valued
ones. Examples are the difference between the density of the liquid
and the density of the gas in a liquid-gas phase transition, the difference in concentrations in binary liquids or the magnetization
in uniaxial magnets. Since the AdS/CFT correspondence is designed
to describe quantum rather than thermal phase transitions on the one hand and strong coupling rather than weak coupling scenarios on the other, the question remains whether such transitions appear in nature. One example is the investigation of liquid-gas quantum phase transitions in bosonic fluids as discussed recently e.g. in \cite{zwerger}. Since $^4$He is known to be a strongly interacting 
system, it would be interesting to compare our results with those obtained using other approaches.

\section{Conclusions}
In this paper, we have studied spontaneously scalarized, static, spherically symmetric (un)charged black holes in asymptotically AdS space-time. We find that the scalar field power law fall-off on the AdS boundary depends on the coupling
constants $\alpha$ and $\gamma$ -- very different to the asymptotically flat case -- and
that this fall-off is possible when the asymptotic form of the scalar-tensor coupling term fulfills he Breitenlohner-Freedman bound. 
We also find that while black holes
can be scalarized with only the Gauss-Bonnet term ${\cal G}$ present, this is not the case for the Ricci scalar ${\cal R}$.
This latter term needs ``the support'' of the Gauss-Bonnet term for scalarization to appear.
Very similar to the asymptotically flat case, RNAdS black holes close to extremality can only be scalarized for negative
values of $\gamma$. Including backreaction of the space-time leads to the observation
that increasing the value of the scalar field on the horizon $\phi_h$ from zero that the branch of solutions terminates
in a singular solution for which $\sigma'$ on the horizon and with it ${\cal R}$ and ${\cal G}$ diverge.
The value of the scalar field on the boundary of AdS can then be interpreted as the expectation value of
a dual operator in a Quantum Field Theory. This value increases with decreasing temperature and the system describes 
a phase transition of the Ginzburg-Landau form. However, since the temperature cannot be decreased to zero, the transition
stops at finite temperature. In order to study temperatures closer to zero, we would have to study solutions
for negative values of $\gamma$. However, our results indicate that in this case, the condensate increases with increasing
temperature. For the moment, we have no proper dual interpretation of this fact, but will report more details including
a detailed study of phase transitions in this system in the future.

\vspace{2cm}

{\bf Acknowledgments} BH would like to thank FAPESP for financial support under grant number 2019/01511-5. 
 NPA thanks CAPES for financial support under grant number 88882.328727/2019-01. 
JU (ORCID ID 0000-0002-4221-2859) acknowledges support from Eusko Jaurlaritza (IT-979-16) and PGC2018-094626-B-C21 (MCIU/AEl/FEDER,UE).

%%%%%%%%%%%%%%%%%%%%%%%%%%%%%%%%%%%%%%%%%%%%%%%%%%%%%%%%%%%%

%%%%%%%%%%%%%%%%%%%%%%%%%%%%%%%%%%%%%%%%%%%%%%%%%%%%%%%%%%%%
\clearpage

\section*{Appendix A: Spontaneous scalarization of extremal black holes}
In the following, we will use the so-called {\it attractor formalism} developed in \cite{Sen:2005wa} for higher deriviative
gravity, in \cite{Astefanesei:2008wz} for higher derivative gravity in AdS and including scalar fields in \cite{Anabalon:2013sra}, respectively.  For that, we assume that the near horizon geometry of a near-extremal spherically symmetric, static black holes is of the form $AdS_2\times S^2$ \footnote{Strictly speaking, this is an assumption. However, we know that for $\phi\equiv 0$ this is true and assume here that the presence of the scalar field would not change the geometry.} and hence can
be written as
\begin{equation}
\label{eq:ads_s2}
ds^2 = v_1 \left(-\rho^2 d\tau^2 + \frac{1}{\rho^2}d\rho^2\right) + v_2 \left(d\theta^2 + \sin^2\theta d\varphi^2\right) \ ,
\end{equation}
where $v_1$ and $v_2$ are two positive constants and $r=r_{h,ex}+\rho$ such that 
the extremal horizon is located at $\rho=0$. Furthermore, we have $F_{\rho\tau}=-F_{\tau\rho}={\rm constant}\equiv e$, which follows from the Maxwell equation using (\ref{eq:ads_s2}).

The entropy function $F$ is given by: $F(v_1,v_2,e,Q,\phi_h)\sim f(v_1,v_2,e,\phi_h)-e Q$, where $f$ reads~:
\begin{equation}
f(v_1,v_2,e,\phi_h)=\int_{S^2} {\rm d}^2 x \sqrt{-g}  {\cal L}
\end{equation}
and 
\begin{equation}
{\cal L}=\left[\frac{{\cal R}}{2} - \Lambda +  \phi^2 \left(\alpha {\cal R} + \gamma {\cal G}\right)  -  \partial_{\mu} \phi \partial^{\mu} \phi  - \frac{1}{4}
 F_{\mu\nu } F^{\mu\nu}  \right]   \ .
\end{equation}
Inserting $F_{\rho\tau}=-F_{\tau\rho}=e$ and the metric (\ref{eq:ads_s2}), we find~:
\begin{equation}
F= 4\pi \left[v_1-v_2       -\frac{\Lambda}{v_1 v_2}  +\phi_h^2\left(2\alpha(v_1-v_2) 
-8\gamma \right) +\frac{1}{2} \frac{e^2 v_2}{v_1}    \right]      -e Q  \ .
\end{equation}
The attractor conditions then read~: 
\begin{eqnarray}
\label{eq:attractor}
\frac{\partial F}{\partial v_1}&=& 0 \ \ \ \ \Longrightarrow  \ \ \ \ 4\alpha\phi_h^2 v_1^2 v_2 - e^2 v_2^2 + 2\Lambda + 2 v_1^2 v_2=0 \nonumber \\
\frac{\partial F}{\partial v_2}&=& 0   \ \ \ \ \Longrightarrow  \ \ \ \ - 4\alpha\phi_h^2 v_1 v_2^2 + e^2 v_2^2 + 2\Lambda - 2 v_1 v_2^2=0 \nonumber \\
\frac{\partial F}{\partial e} &=& 0  \ \ \ \ \Longrightarrow  \ \ \ \ 4\pi  e \frac{v_2}{v_1} = Q  \nonumber \\
\frac{\partial F}{\partial \phi_h} &=& 0  \ \ \ \ \Longrightarrow  \ \ \ \ \phi_h\left(\alpha(v_1 - v_2) - 4\gamma\right)=0
\end{eqnarray}
The latter expression clearly demonstrates that for $\phi_h\neq 0$ the choice $\alpha=0$ implies $\gamma=0$ (and vice versa) and hence
scalarization of black holes with near horizon geometry given by $AdS_2\times S^2$ is excluded
if either term is absent, unless $v_1=v_2$ in the case $\alpha\neq 0$, $\gamma=0$. Setting $v_1=v_2$, however,
leads to $\Lambda=0$ (using the first two relations in (\ref{eq:attractor})), a case we want to exclude here.  

Now, for the general case, $\alpha\neq 0$, $\gamma\neq 0$ and $\phi_h\neq 0$, let us consider the scalar field equation.
Using the metric (\ref{eq:ads_s2}), this reads~:
\begin{equation}
\label{scalareq}
8\gamma \phi - 2\alpha \phi (v_1 - v_2) - 2\phi' v_2 \rho
- \phi'' v_2 \rho^2 = 0  \ .
\end{equation}
Implementing the condition $\alpha(v_1 - v_2) = 4\gamma$ from (\ref{eq:attractor}), this equation can be integrated to give
\begin{equation}
\phi' \sim \rho^{-2}    \ .
\end{equation}
Hence, $\phi'$ diverges when approaching the extremal horizon, i.e. for $
\rho\rightarrow 0$, such that {\it regular, extremal black holes with scalar hair do not exist in scalar-tensor gravity with coupling of the form $\phi^2\left(\alpha {\cal R}+\gamma {\cal G}\right)$}.

\section*{Appendix B: Fefferman-Graham construction}
In the Fefferman-Graham construction \cite{fg} the bulk metric is of the form~:
\begin{equation}
\label{eq:FG}
{\rm d}s^2 = \frac{\ell^2}{4\rho^2} {\rm d} \rho^2 + \frac{1}{\rho} g_{ij}(x,\rho) {\rm d}x^i {\rm d}x^j   \ ,
\end{equation}
where $\rho$ is the radial coordinate that drives the RG flow in the holographic interpretation. 
Anti-de Sitter space-time (AdS) corresponds to the choice
\begin{equation}
\label{eq:adsFG}
g_{ij}(x,\rho) {\rm d}x^i {\rm d}x^j = -{\rm d}t^2 + {\rm d}x^2 + {\rm d}y^2   \ ,
\end{equation}
where $x,y,z$ are standard Cartesian coordinates, and is a solution to the Einstein equation with negative
cosmological constant $\Lambda=-\frac{3}{\ell^2}$. Since we are dealing with a scalar-tensor gravity model
here, we have checked that the space-time, indeed, tends to AdS asymptotically. For that, we have chosen 
\begin{equation}
 g_{ij}(x,\rho) = -\left(1+a(\rho)\right)dt^2 + \left(1+b(\rho)\right) \left(dx^2 + dy^2\right)  \ \ , \ \  \text{with} \ \ \ 
 a(\rho)=\sum\limits_{k=1}^{\infty} a_k \rho^k \ \ , \ \   b(\rho)=\sum\limits_{k=1}^{\infty}  b_k \rho^k 
\end{equation}
where the $a_k$ and $b_k$ are constants. This
seems a suitable Ansatz given the symmetries of the space-time. Moreover, we assume $\phi \sim \rho^{\beta}$ with $\beta \geq 1 $ at $\rho\rightarrow 0$. 
Inserting this into the gravity equation (see rhs of (\ref{eq:scalar})) it is straightforward to show that the
dominant term in $T_{\mu\nu}^{(\phi)}$ is the scalar field
term $\sim \rho^{2(\beta-1)}$. Moreover, this can never 
be matched with the terms from $a(\rho)$ and
$b(\rho)$ appearing in $G_{\mu\nu}$. Hence, we conclude
that $a_k=0$ and $b_k=0$ for all $k$ when $\rho\rightarrow 0$
and that the space-time does tend to ``pure'' AdS, i.e. (\ref{eq:adsFG}), in our scalar-tensor gravity model. Note that this
is also in agreement with our numerical results. 
Now, using this fact, it is straightfoward to show using the scalar 
field equation (lhs of (\ref{eq:scalar})) that the scalar field behaves like (\ref{eq:behaviour_boundary}) with $r$ replaced by $\rho^{-1}$  for $
\rho\rightarrow 0$. This means that the power of the fall-off that we have derived
in Schwarzschild-like coordinates agrees with that in the Fefferman-Graham construction.


\begin{thebibliography}{99}
%%%%%%%%%%%%%%%%%%%%%%%%%%%%%%%%%%%%%%%%%%%%%%%%%%%%%%%%%%%% 

\bibitem{LIGO}  
B.~P.~Abbott {\it et al.} [LIGO Scientific and Virgo Collaborations], {\it Observation of Gravitational Waves from a Binary Black Hole Merger},
  Phys.\ Rev.\ Lett.\  {\bf 116} (2016) no.6,  061102; {\it 
  GW170817: Observation of Gravitational Waves from a Binary Neutron Star Inspiral},
  Phys.\ Rev.\ Lett.\  {\bf 119} (2017) no.16.
\bibitem{EHT} K.~Akiyama {\it et al.} [Event Horizon Telescope Collaboration], {\it First M87 Event Horizon Telescope Results. I. The Shadow of the Supermassive Black Hole}, Astrophys.\ J.\  {\bf 875} (2019) no.1,  L1. 
\bibitem{CHANDRA}  E.~Troja {\it et al.}, {\it The X-ray counterpart to the gravitational wave event GW 170817},
  Nature {\bf 551} (2017) 71.
\bibitem{nohairtheorems}
  W.~Israel, {\it Event horizons in static vacuum space-times},
  Phys.\ Rev.\  {\bf 164} (1967) 1776;
  S.~W.~Hawking, {\it Black holes in general relativity}, 
  Commun.\ Math.\ Phys.\  {\bf 25} (1972) 152; B. Carter, {\it Axisymmetric Black Hole Has Only Two Degrees of Freedom}, Phys. Rev. Lett. {\bf 26}, 331 (1971);
{\it The vacuum black hole uniqueness theorem and its conceivable generalisations}, Proceedings of the 1st Marcel Grossmann meeting 
on general relativity, 243 (1977); D. Robinson, {\it Uniqueness of the Kerr Black Hole}, Phys. Rev. Lett. {\bf 34}, 905 (1975) ; M. Heusler, {\it Stationary Black Holes: Uniqueness and Beyond}, Liv. Rev. Rel. 1 (1998); C.~Misner, K.~Thorne, and J.~Wheeler, {\it Gravitation}, W. H. Freeman and Company, (1973).


\bibitem{AGK}  A.~Achucarro, R.~Gregory and K.~Kuijken, {\it Abelian Higgs hair for black holes}, 
  Phys.\ Rev.\ D {\bf 52} (1995) 5729.
  \bibitem{GKW}  R.~Gregory, D.~Kubiznak and D.~Wills, {\it Rotating black hole hair}, 
  JHEP {\bf 1306} (2013) 023. 

\bibitem{hairy_BHs} {\it see e.g.} H.~Luckock and I.~Moss, {\it Black Holes Have Skyrmion Hair}, 
  Phys.\ Lett.\ B {\bf 176} (1986) 341; K.~Lee, V.~P.~Nair and E.~J.~Weinberg, {\it Black holes 
  in magnetic monopoles}, Phys. Rev. D {\bf 45} (1992) 2751; P.~Breitenlohner, P.~Forgacs and D.~Maison, {\it Gravitating monopole solutions}, 
Nucl. Phys. B {\bf 383} (1992) 357;
{\it Gravitating monopole solutions II}, Nucl. Phys. B {\bf 442} (1995) 126;
P.~C.~Aichelburg and P.~Bizon, {\it Magnetically charged black holes and their stability}, Phys. Rev. D {\bf 48} (1993) 607. 



%%%%%%%%%%%%%%%

%%%% scalar fields + BH %%%%%%%%%%%%%

\bibitem{no_scalar_hair} J.~E.~Chase,  {\it Event horizons in static scalar-vacuum space-times},  Commun.~Math.~Phys. {\bf 19} (1970) 276;
  J.~D.~Bekenstein, {\it Transcendence of the law of baryon-number conservation in black hole physics},  Phys.\ Rev.\ Lett.\  {\bf 28} (1972) 452; C. ~Teitelboim, Lett. Nuovo Cimento {\bf 3} (1972) 326;  J.~D.~Bekenstein, {\it Novel "No hair theorem" for black holes}, Phys. Rev. D {\bf 51} (1992), R6608.

%%%% Horndeski et al %%%%

[18] G. W. Horndeski,Second-order scalar-tensor field equations in a four-dimensional space, Int. J.Theor. Phys.10, 363 (1974).[19] C. Deffayet and D. A. Steer,A formal introduction to Horndeski and Galileon theories and theirgeneralizations, Class. Quant. Grav.30, 214006 (2013).[20] C. Charmousis,From Lovelock to Horndeski‘s Generalized Scalar Tensor Theory, Lect. Notes Phys.892(2015) 25.[21] T. P. Sotiriou and S. Y. Zhou,Black hole hair in generalized scalar-tensor gravity: An explicitexample, Phys. Rev. D90, 124063 (2014)[22] E. Babichev, C. Charmousis and A. Leh ́ebel,Asymptotically flat black holes in Horndeski theoryand beyond, JCAP1704(2017), 027.

%%%% Horndeski scalar-tensor

 \bibitem{horndeski} G. W. Horndeski,  	
{\it Second-order scalar-tensor field equations in a four-dimensional space}, Int. J. Theor. Phys. {\bf 10}, 363 (1974).


\bibitem{Deffayet:2013lga}
  C.~Deffayet and D.~A.~Steer, {\it A formal introduction to Horndeski and Galileon theories and their generalizations},
  Class.\ Quant.\ Grav.\  {\bf 30}, 214006 (2013).

  \bibitem{Charmousis:2014mia}
  C.~Charmousis, {\it From Lovelock to Horndeski`s Generalized Scalar Tensor Theory},
  Lect.\ Notes Phys.\  {\bf 892} (2015) 25.
  

\bibitem{Sotiriou:2014pfa}
  T.~P.~Sotiriou and S.~Y.~Zhou, {\it Black hole hair in generalized scalar-tensor gravity: An explicit example},
  Phys.\ Rev.\ D {\bf 90}, 124063 (2014) 
  
\bibitem{Babichev:2017ab}  E.~Babichev, C.~Charmousis and A.~Leh\'ebel,
{\it Asymptotically flat black holes in Horndeski theory and beyond},  JCAP {\bf 1704} (2017), 027.



%%Spontaneous scalarization%%%

\bibitem{Silva:2017uqg}
  H.~O.~Silva, J.~Sakstein, L.~Gualtieri, T.~P.~Sotiriou and E.~Berti, {\it Spontaneous scalarization of black holes and compact stars from a Gauss-Bonnet coupling}, 
  Phys.\ Rev.\ Lett.\  {\bf 120} (2018) ,  131104.

\bibitem{Doneva:2017bvd}
  D.~D.~Doneva and S.~S.~Yazadjiev, {\it New Gauss-Bonnet Black Holes with Curvature-Induced Scalarization in Extended Scalar-Tensor Theories}, 
  Phys.\ Rev.\ Lett.\  {\bf 120} (2018) ,  131103.

  
  
\bibitem{Antoniou:2017acq}
  G.~Antoniou, A.~Bakopoulos and P.~Kanti, {\it Evasion of No-Hair Theorems and Novel Black-Hole Solutions in Gauss-Bonnet Theories}, 
  Phys.\ Rev.\ Lett.\  {\bf 120} (2018),  131102

  
\bibitem{Antoniou:2017hxj}
  G.~Antoniou, A.~Bakopoulos and P.~Kanti, {\it Black-Hole Solutions with Scalar Hair in Einstein-Scalar-Gauss-Bonnet Theories}, 
  Phys.\ Rev.\ D {\bf 97} (2018),  084037.

\bibitem{Minamitsuji:2018xde}
  M.~Minamitsuji and T.~Ikeda, {\it Scalarized black holes in the presence of the coupling to Gauss-Bonnet gravity}, 
  Phys.\ Rev.\ D {\bf 99} (2019) 044017.
  
  
  \bibitem{Brihaye:2018grv}   Y.~Brihaye and L.~Ducobu,  {\it Hairy black holes, boson stars and non-minimal coupling to curvature invariants}, Phys.\ Lett.\ B {\bf 795} (2019) 135. 
  
 
  
  %%%% + electric fields %%%%
  
\bibitem{Doneva:2018rou}
  D.~D.~Doneva, S.~Kiorpelidi, P.~G.~Nedkova, E.~Papantonopoulos and S.~S.~Yazadjiev,
{\it Charged Gauss-Bonnet black holes with curvature induced scalarization in the extended scalar-tensor theories}, 
  Phys.\ Rev.\ D {\bf 98} (2018),  104056. 
  
  \bibitem{Herdeiro:2018wub}
  C.~A.~R.~Herdeiro, E.~Radu, N.~Sanchis-Gual and J.~A.~Font, {\it Spontaneous Scalarization of Charged Black Holes}, 
  Phys.\ Rev.\ Lett.\  {\bf 121} (2018) no.10,  101102.
  
  \bibitem{Brihaye:2019kvj}
  Y.~Brihaye and B.~Hartmann, {\it Spontaneous scalarization of charged black holes at the approach to extremality}, 
  Phys.\ Lett.\ B {\bf 792} (2019) 244. 
  
\bibitem{Bakopoulos:2018nui}
  A.~Bakopoulos, G.~Antoniou and P.~Kanti, {\it Novel Black-Hole Solutions in Einstein-Scalar-Gauss-Bonnet Theories with a Cosmological Constant}, 
  Phys.\ Rev.\ D {\bf 99} (2019) no.6,  064003. 
  
    %\cite{Brihaye:2019gla} 
  \bibitem{Brihaye:2019gla}  
   Y.~Brihaye, C.~Herdeiro and E.~Radu,  {\it Black hole spontaneous scalarisation with a positive cosmological constant}, 
  arXiv:1910.05286 [gr-qc].
  
%\cite{Bakopoulos:2019tvc} 
\bibitem{Bakopoulos:2019tvc}   
A.~Bakopoulos, P.~Kanti and N.~Pappas,  
{\it On the Existence of Solutions with a Horizon in Pure Scalar-Gauss-Bonnet Theories}, 
  arXiv:1910.14637 [hep-th].  
  
  
  
  


  
 %%%%% AdS/CFT %%%%%%%%%


 
\bibitem{ggdual} 
{\it see e.g.} O.~Aharony, S.~S.~Gubser, J.~M.~Maldacena, H.~Ooguri and Y.~Oz, {\it Large N field theories, string theory and gravity},
  Phys.\ Rept.\  {\bf 323} (2000) 183
  [arXiv:hep-th/9905111];
 E.~D'Hoker and D.~Z.~Freedman, {\it Supersymmetric gauge theories and the AdS/CFT correspondence}, 
  arXiv:hep-th/0201253;
M. Benna and I. Klebanov, {\it Gauge-string duality and some applications} [arXiv: 0803.1315 [hep-th]].   

\bibitem{adscft} J. Maldacena, Adv. Theo. Math. Phys. {\bf 2} (1998) 231;
Int. J. Theor. Phys. {\bf 38} (1999) 1113.

\bibitem{Gubser:2008px}
  S.~S.~Gubser, {\it Breaking an Abelian gauge symmetry near a black hole horizon}, 
  Phys.\ Rev.\ D {\bf 78} (2008) 065034.  
  
 \bibitem{hhh} S.~A.~Hartnoll, C.~P.~Herzog and G.~T.~Horowitz, {\it Holographic Superconductors},
  JHEP {\bf 0812} (2008) 015; {\it Building a Holographic Superconductor},
  Phys.\ Rev.\ Lett.\  {\bf 101} (2008) 031601.
\bibitem{reviews} {\it for reviews see} C.~P.~Herzog, {\it Lectures on Holographic Superfluidity and Superconductivity}, 
  J.\ Phys.\  {\bf A42} (2009) 343001;
  S.~A.~Hartnoll, {\it Lectures on holographic methods for condensed matter physics},
  Class.\ Quant.\ Grav.\  {\bf 26} (2009) 224002;
G. Horowitz, {\it Introduction to holographic superconductors},
arXiv:1002.1722 [hep-th]; S.~A.~Hartnoll, {\it Horizons, holography and condensed matter},
  arXiv:1106.4324 [hep-th].
 
\bibitem{Ammon:2015wua}
  M.~Ammon and J.~Erdmenger, {\it Gauge/gravity duality : Foundations and applications}, Cambridge University Press (2015).
  
  \bibitem{BF}  P.~Breitenlohner and D.~Z.~Freedman, {\it Stability in Gauged Extended Supergravity}, 
  Annals Phys.\  {\bf 144} (1982) 249.

  
%%%% AdS/CFT interpretation
\bibitem{Skenderis:2002wp}
  K.~Skenderis, {\it Lecture notes on holographic renormalization}, 
  Class.\ Quant.\ Grav.\  {\bf 19} (2002) 5849.




%%%% else

\bibitem{COLSYS}
 U. Ascher, J. Christiansen, R.~D. Russell,
% A collocation solver for mixed order systems of boundary value problems,
 Math. of Comp. {\bf 33} (1979) 659;\\
 U. Ascher, J. Christiansen, R.~D. Russell,
%Collocation software for boundary-value ODEs,
 ACM Trans. {\bf 7} (1981) 209.
 
 
 \bibitem{Sen:2005wa}
  A.~Sen, {\it Black hole entropy function and the attractor mechanism in higher derivative gravity}, 
  JHEP {\bf 0509} (2005) 038.
  
  \bibitem{Astefanesei:2008wz}
  D.~Astefanesei, N.~Banerjee and S.~Dutta, {\it (Un)attractor black holes in higher derivative AdS gravity}, 
  JHEP {\bf 0811} (2008) 070.
  
  \bibitem{Anabalon:2013sra}
  A.~Anabal\'on and D.~Astefanesei, {\it On attractor mechanism of $AdS_{4}$ black holes}, 
  Phys.\ Lett.\ B {\bf 727} (2013) 568. 
  
  \bibitem{zwerger} W. Zwerger, {\it Quantum unbinding near a zero temperature liquid-gas transition}, J. Stat. Mech. (2019) 103104
  
  \bibitem{fg} {\it for a review see} 
C.~Fefferman and C.~R.~Graham, {\it The ambient metric}
Ann. Math. Stud. \textbf{178} (2011) 1.

%%%%%%%%%%%%%%%%%%%%%%%%%%%%%%%%%%%%%%%%%%%%%%%%%%%%%%%%%%%% 
%%%%%%%%%%%%%%%%%%%%%%%%%%%%%%%%%%%%%%%%%%%%%%%%%%%%%%%%%%%% 

\end{thebibliography}
 \end{document}